# Multiphasic profiles for voltage-dependent K$^+$ channels: Reanalysis of data of MacKinnon and coworkers

Per Nissen

Norwegian University of Life Sciences
Department of Ecology and Natural Resource Management
P. O. Box 5003, NO-1432 Ås, Norway

per.nissen@nmbu.no

2016

# Abstract

In a study of the role that voltage-dependent $K^+$ channels may have in the mechanosensation of living cells (Schmidt et al. *Proc Soc Natl Acad Sci USA* 109: 10352-10357. 2012), the data were as conventionally done fitted by a Boltzmann function. However, as also found for other data for ion channels, this interpretation must be rejected in favor of a multiphasic profile, a series of straight lines separated by discontinuous transitions, quite often in the form of noncontiguities (jumps). The data points in the present study are often very unevenly distributed around the curvilinear profiles. Thus, for 43 of the 75 profiles, the probability is less than 5% that the uneven distribution is due to chance, for 26 the probability is less than 1%, and for 12 the probability is less than 0.1%, giving a vanishingly low overall probability for all profiles. Especially at low voltages, the differences between the fits to curvilinear and multiphasic profiles may be huge. In the multiphasic profiles, adjacent lines are quite often parallel or nearly so, in which case the transitions are, necessarily, in the form of jumps. The lines in the multiphasic profiles appear to be perfectly straight, with no indication of any curvilinearity. The r values are for the most part high, quite often exceedingly high.

In addition to activation of ion channels, a wide variety of biological as well as non-biological processes and phenomena involving binding, pH, folding/unfolding and effect of chain length can be well represented by multiphasic profiles (Nissen 2015a,b, 2016. Posted on arXiv.org with Paper ID arXiv: 1511.06601, 1512.02561 and 1603.05144).

# Introduction

In addition to multiphasic profiles for ion uptake in plants (Nissen 1971, 1974, 1991, 1996), such profiles have been recently (Nissen 2015a,b, Nissen 2016) reported for many other processes and phenomena. In the present paper, data (Schmidt et al. 2012) for voltage-dependent Kv channels will be reanalyzed to compare the fits to curvilinear profiles with the fits to multiphasic profiles.

Original data have been kindly provided by Daniel Schmidt. In addition to the r values, slopes ± SE (or only slopes) are given on the plots. Except for the data for Figs 11-13 and 27-29 (authors' Figs 1L and S2C), all slopes have been multiplied by 1000. The Runs test (Wald and Wolfowitz 1940) gives the probability for the uneven distribution of points around the curvilinear profile being due to chance. Independent probabilities have been combined by the method of Fisher (Fisher 1954).

# Reanalysis

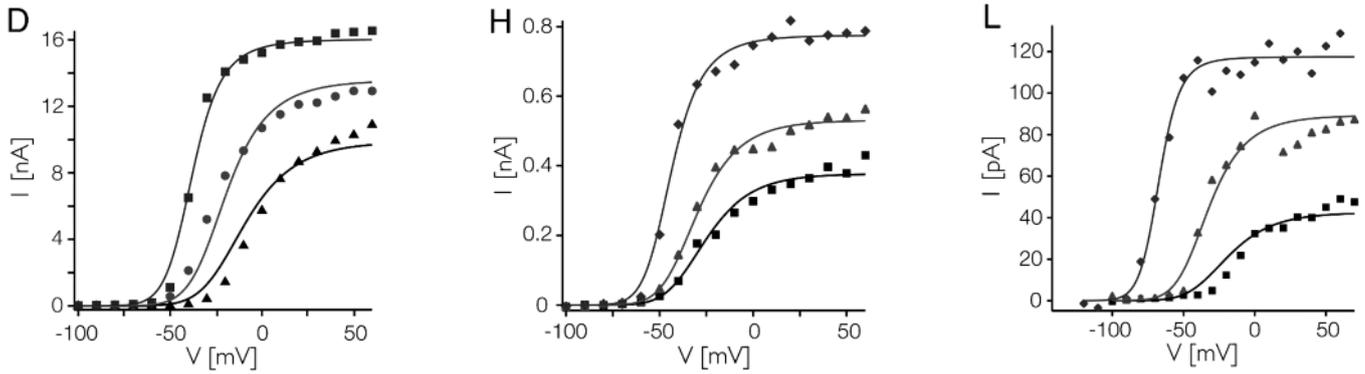

**Fig. 1.** Authors' Fig. 1D, H and L. Plot D: Paddle chimera in the same outside-out patch with 0 mm Hg (triangles), 5 mm Hg (circles) and 15 mm Hg (squares) of transient pressure applied. Plot H: Shaker Kv in the same outside-out patch at different time points after patch excision: 0 min (squares), 4 min (triangles) and 8 min (diamonds). Plot L: Kv2.1 channels in the same on-cell patch with 0 mm Hg (squares), 5 mm Hg (triangles) and 15 mm Hg (diamonds) of transient pressure applied. See also the original legend.

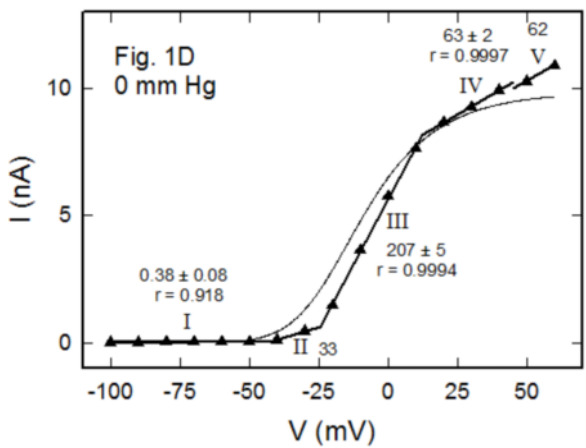

Fig. 2. Reanalysis of data in Fig. 1. The profile for 0 mm Hg can be well represented as pentaphasic, with the transitions at -42.6, -24.3 and 12.2, and between 40 and 50 mV (jump). The r values for lines III and IV are very high and lines IV and V are precisely parallel, indicating that the data are highly precise. The 6-point line I has a low r value. This could, most likely, be because of its very shallow slope (slight errors will cause marked decreases in the r values for such lines). However, there could also be several phases in this range (the line for -90 and -80 mV and the line for -60 and -50 mV have about the same slope, 1.0 and 1.1, respectively).

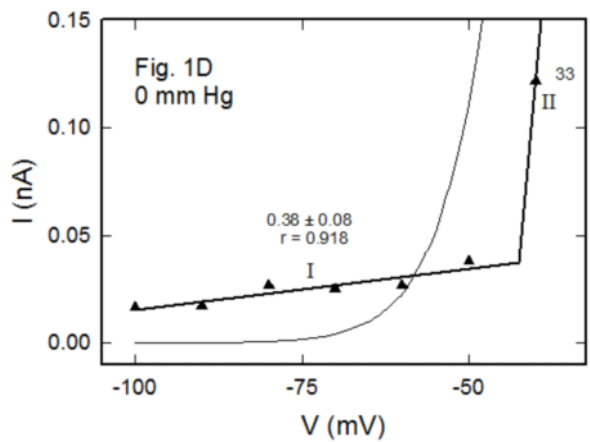

**Fig. 3.** The enlargement of Fig. 2 at high negative voltages shows the huge differences in the fits to the curvilinear and the multiphasic profile.

In contrast to the precisely multiphasic profiles, the fits to the curvilinear profile are poor. Furthermore, the points are very unevenly distributed around this profile. Points 1-5 and 13-17 are above the line, points 6-12 are below the line. The probability of this uneven distribution, or an even more unlikely one, occurring by chance is only 0.09% (by the Runs test).

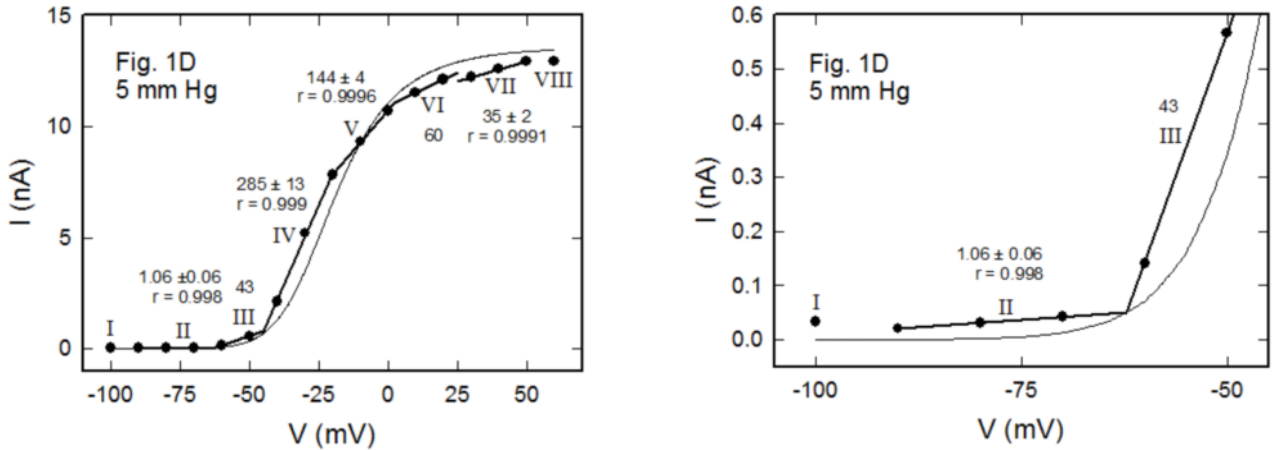

**Fig. 4** (left above). Reanalysis of data in Fig. 1. The profile for 5 mm Hg can be well represented by 8 phases, with the transitions between -100 and -90, at -62.2, -45.0, -20 and 2.3, between 20 and 30 (jump), and between 50 and 60 mV. The r values for lines II, IV, V and VII are very high. The multiphasic profile is precise, and the curvilinear profile is imprecise.

**Fig. 5** (right above). Although not evident from Fig. 4, this enlargement shows that the fits to the curvilinear profile are very poor also at the lowest voltages. (A single line in the range of phases I and II will have an r value of only 0.547.)

The fits to the curvilinear profile are poor when compared to the multiphasic profile. Furthermore, the distribution around the curvilinear profile is exceedingly uneven. The first 10 points are all above the line, the remaining 7 are all below the line. The probability of this happening by chance is only 0.01%.

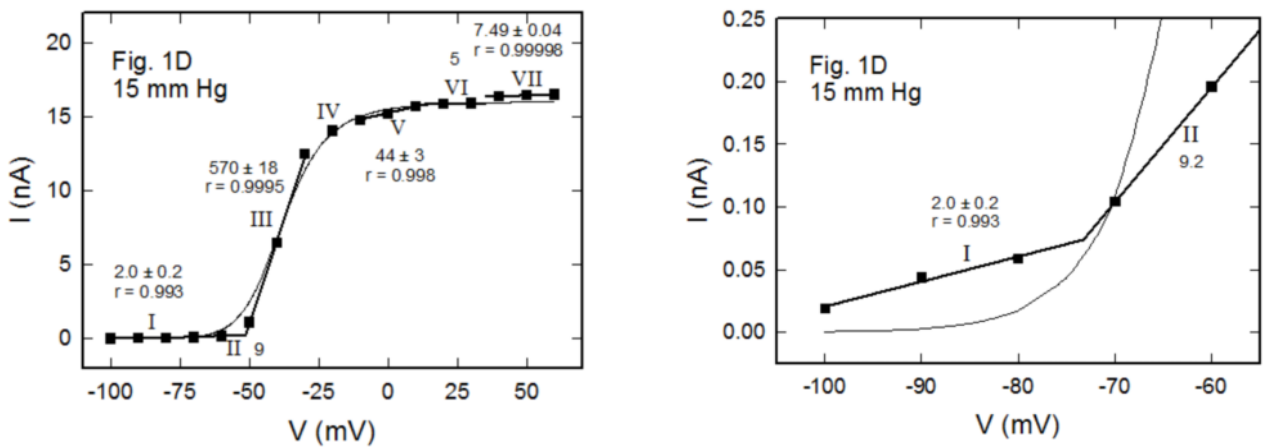

**Fig. 6** (left above). Reanalysis of data in Fig. 1. The profile for 15 mm Hg can be well represented as heptaphasic, with the transitions at -73.3 and -51.3, between -30 and -20, between -20 and -10, at 13.4, and between 30 and 40 mV (jump). The data are insufficiently detailed in the range of phase IV for the line to be resolved. The r values are high for lines I and V, and very to exceedingly high for lines III and VII.

**Fig. 7** (right above). As also for lower Hg (Figs 3 and 5), the data can be well represented as multiphasic, but not as curvilinear, also at low y values (right), where the curvilinear profile gives, for the most part, exceedingly poor fits.

The distribution of points around the curvilinear profile is somewhat uneven. Points 1-3, 8-9, 13 and 15-17 are above the profile, points 4-7, 10-12, and 14 are below the curvilinear line. The probability of this or a more uneven distribution being due to chance is 15.7%. The probability that the uneven distribution around the three profiles in Fig. 1D is due to chance is less than 0.001% (by Fisher's method for combining independent probabilities).

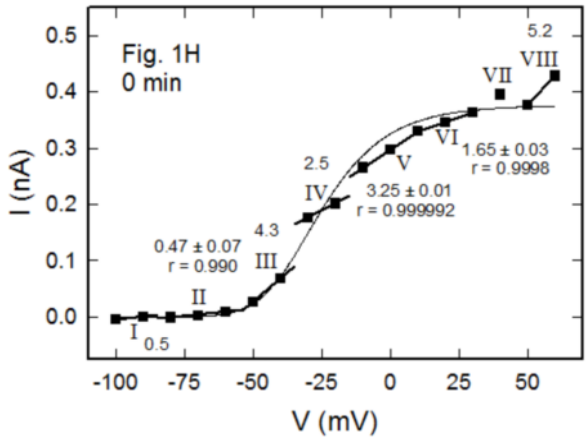

**Fig. 8.** Reanalysis of data in Fig. 1H. The profile for 0 min can be well represented by 8 phases, with the transitions between -90 and -80 (jump), at -53.4, between -40 and -30 (jump), between -20 and -10 (jump), at 10, between 30 and 40, and between 40 and 50 mV. The data are insufficiently detailed in the range of phase VII for the line to be resolved. Lines I and II are parallel. The r values for lines V and VI are very to exceedingly high. Runs test: P = 31.9%.

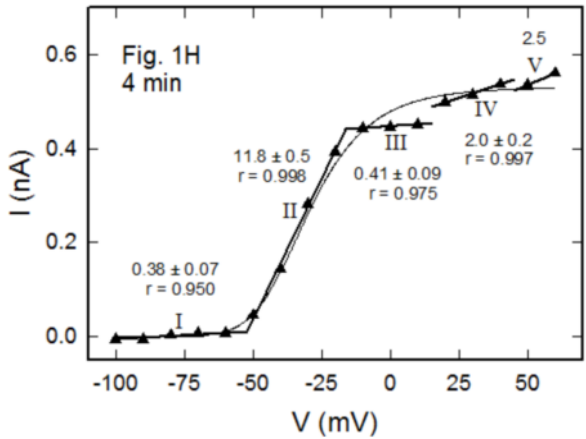

**Fig. 9.** The profile for 4 min can be well represented as pentaphasic, with the transitions at -52.2 and -16.0, between 10 and 20 (jump), and between 40 and 50 mV (jump). Lines I and III are parallel, lines IV and V are about parallel. Runs test: P = 6.9%.

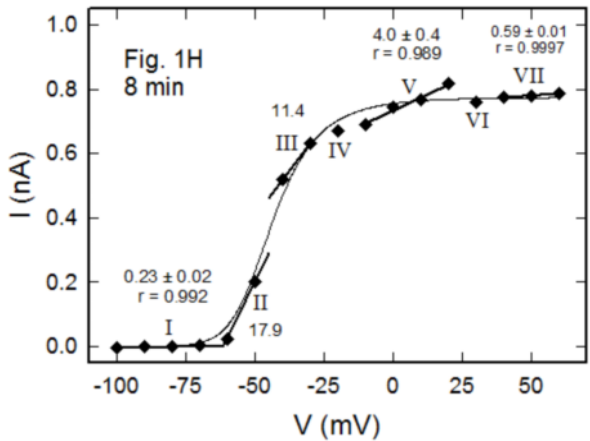

**Fig. 10.** The profile for 8 min can be well represented as heptaphasic, with the transitions at -61.0, between -50 and -40 (jump), between -30 and -20, between -20 and -10, between 20 and 30, and between 30 and 40 mV. The data are insufficiently detailed in the range of phases IV and VI for the lines to be resolved. The r value for line VII is high. Runs test: P = 8.0%.

The probability that the uneven distribution around the three profiles in Fig. 1H is due to chance is less than 5%.

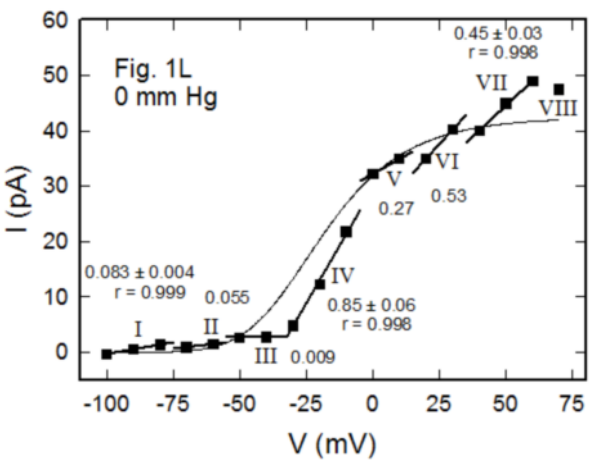

**Fig. 11.** Reanalysis of data in Fig. 1L. The profile for 0 mm Hg can be well represented by 8 phases, with the transitions between -80 and -70 (jump), between -60 and -50 (jump), at -31.9 mV, between -10 and 0 (jump), between 10 and 20 (jump), between 30 and 40 (jump), and between 60 and 70 mV. A single line in the range of lines I-III will have an r value of only 0.941. Lines VI and VII are about parallel. Runs test: P = 23.8%.

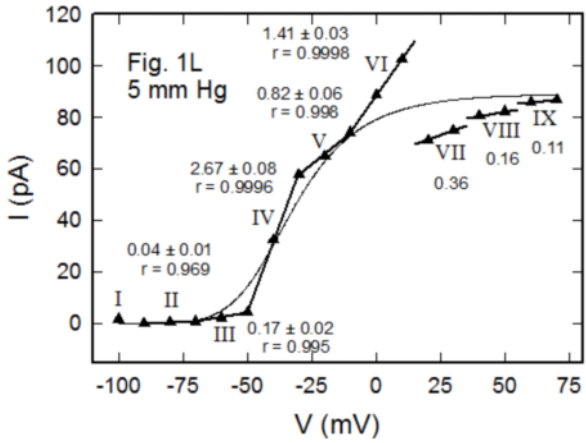

**Fig. 12.** The profile for 5 mm Hg can be well represented by 9 phases, with the transitions between -100 and -90, at -70, -50, -30 and -10, between 10 and 20 (jump), between 30 and 40 (jump), and between 50 and 60 mV (jump). Line II is essentially horizontal. A single line in the range of lines VII-IX will have an r value of 0.979, which is low compared to the r values for lines III-VI. Lines VIII and IX are about parallel. Runs test: P = 0.30%.

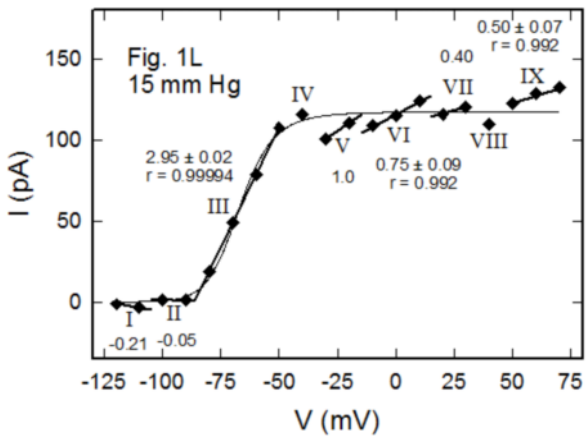

**Fig. 13.** The profile for 15 mm Hg can be well represented by 9 phases, with the transitions between -110 and -100 , at -86.2 , between -50 and -40, between -40 and -30, between -20 and -10 (jump), between 10 and 20 (jump), between 30 and, and between 40 and 50 mV. The data are insufficiently detailed in the range of phases IV and VIII for the lines to be resolved. Line II has a very high r value. Lines V and VI are about parallel, as are lines VII and IX. Runs test: P = 75.8%.

The probability that the uneven distribution around the three profiles in Fig. 1L is due to chance is less than 2.5%.

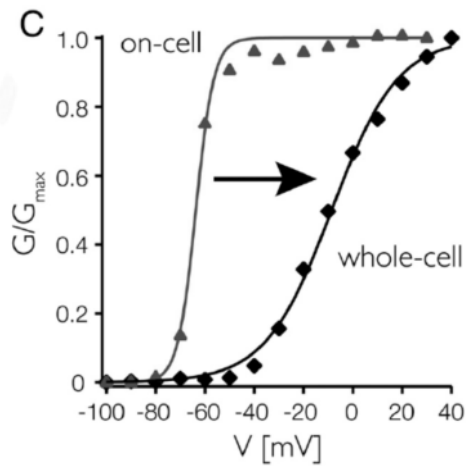

**Fig. 14.** Authors' Fig. 2C. Paddle chimera in on-cell patch (see authors' Fig. 2D) or whole-cell configuration (see authors' Fig. 2E) recorded from the same Sf-9 cell. Solid lines, Boltzmann functions. (But see reanalysis on next page.)

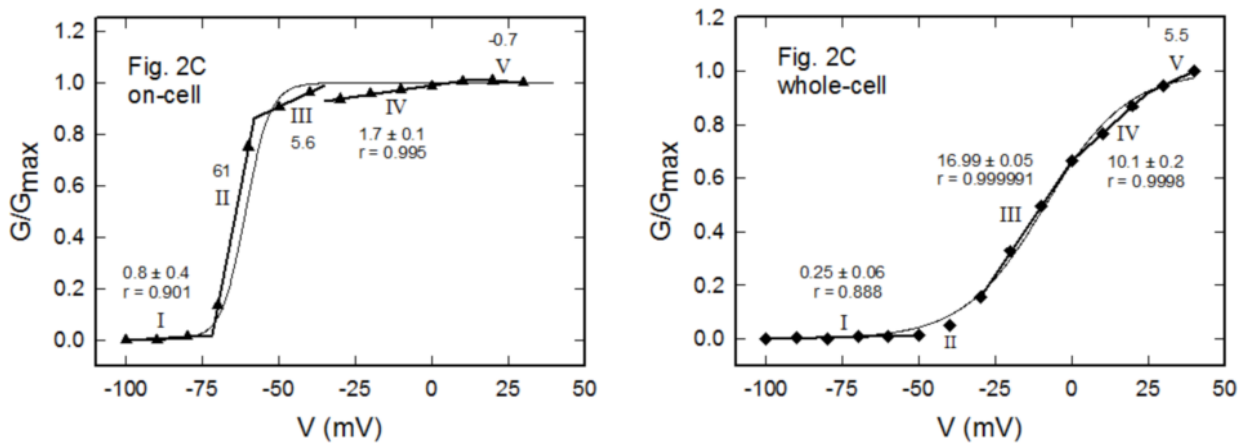

**Fig. 15** (left above). Reanalysis of data in Fig. 14. The profile for the on-cell data can be well represented as pentaphasic, with the transitions at -71.9 and -58.2, between -40 and -30 (jump), and at 13.8 mV. Runs test: P = 2.51%.

**Fig. 16** (right above). Reanalysis of data in Fig. 14. The profile for the whole-cell data can also be well represented as pentaphasic, with the transitions between -50 and -40 mV, between -40 and -30 mV, and at 0 and 24.8 mV. The r value is very high for line IV, and exceedingly high for line III. Runs test: P = 56.6%.

The probability that the uneven distribution around the two profiles in Fig. 14 is due to chance is less than 10%.

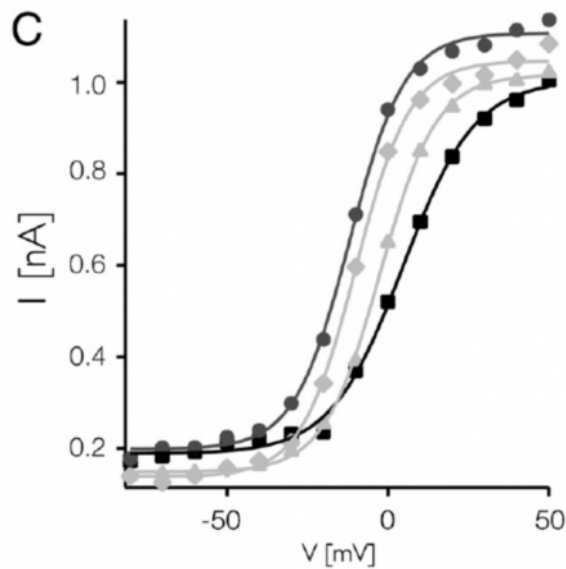

**Fig. 17.** Authors' Fig. 4C. The effect of swelling by hypoosmotic shock. Boltzmann functions for Paddle chimera expressed in sF-9 cells before swelling (squares), at intermediate swelling state 1 (triangles, at intermediate swelling state 2 (diamonds), and at peak volume (circles). See also the original legend and reanalyses on the next page.

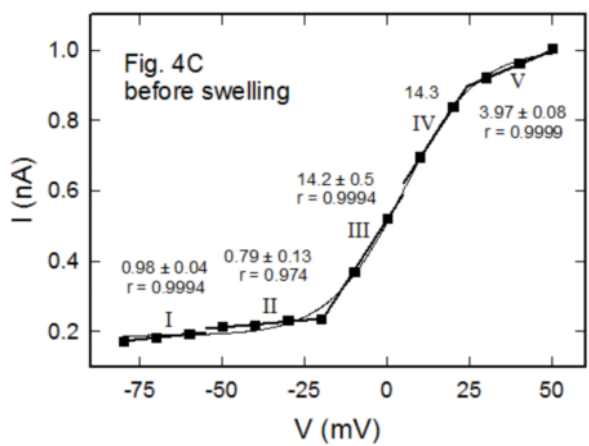 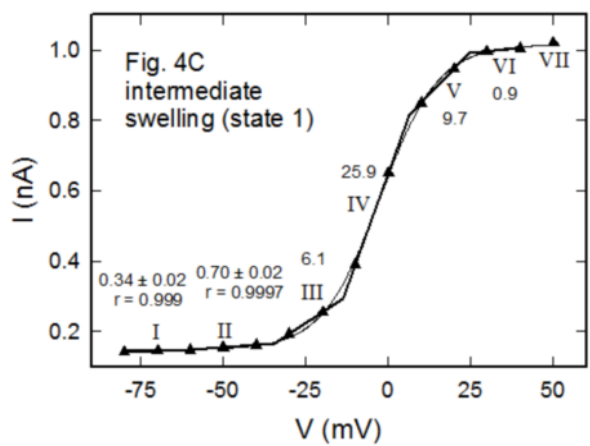

**Fig. 18** (left above). Reanalysis of data in Fig. 17. The profile before swelling can be well represented as pentaphasic, with the transitions between -60 and -50 (jump), at -20, between 0 and 10 (jump), and at 24.2 mV. Lines I and II are parallel, as are lines III and IV. The r values for lines I, III and V are very high. Runs test: P = 22.6%.

**Fig. 19** (right above). The profile for intermediate swelling state 1 can be well represented by 7 phases, with the transitions at -60, -34.5, -13.8, 6.3 and 24.6, and between 40 and 50 mV. The r values are high for lines I and II. Runs test: P = 64.6%.

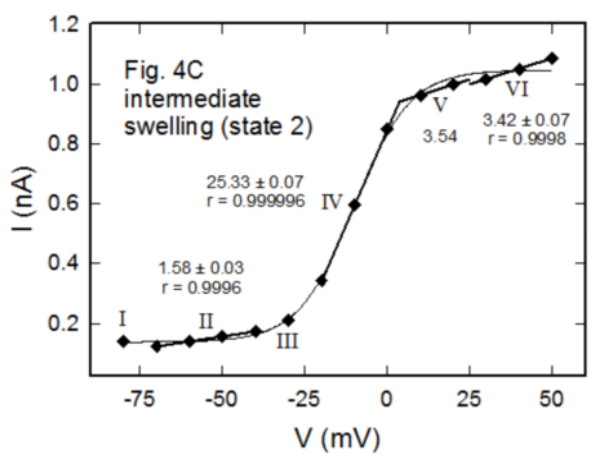 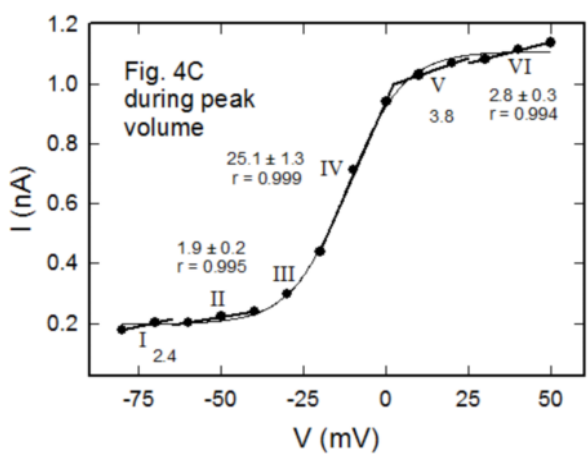

**Fig. 20** (left above). The profile for intermediate swelling state 2 can be well represented as hexaphasic, with the transitions between -80 and -70, between -40 and -30, between -30 and -20, at 3.6, and between 20 and 30 mV (jump). The data are insufficiently detailed in the range of phase III for the line to be resolved. The r values for lines II and VI are very high, that for line IV is exceedingly high. Lines V and VI are parallel. Runs test: P = 38.3%.

**Fig. 21** (right above). The profile during peak volume can also be well represented as hexaphasic, with the transitions between -70 and -60 (jump), between -40 and -30, between -30 and -20, at 2.1, and between 20 and 30 mV (jump). The data are insufficiently detailed in the range of phase III for the line to be resolved. Lines V and VI are about parallel. Runs test: P = 28.7%.

The probability that the uneven distribution around the four profiles in Fig. 17 is due to chance is somewhat more than 10%.

The profiles for the two independent experiments in Figs 20 and 21 have remarkably similar multiphasic patterns, with only a slight difference at high negative voltages. The essentially identical patterns, including the two parallel lines V and VI, constitute conclusive evidence that the patterns are indeed multiphasic. (Clearly, two so strikingly similar patterns cannot have arisen by chance.)

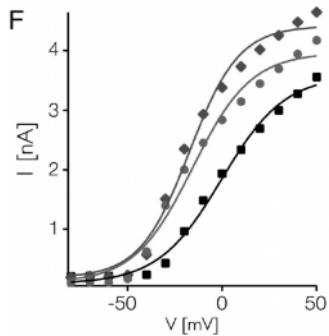

**Fig. 22.** Authors' Fig. 4F. The effect of swelling by hypoosmotic shock. Boltzmann functions for Paddle chimera expressed in Sf-9 cells during consecutive perfusion with iso-osmotic solution (squares), hypoosmotic solution (circles), and immediately after peaking volume (diamonds). See also the original legend and reanalyses below.

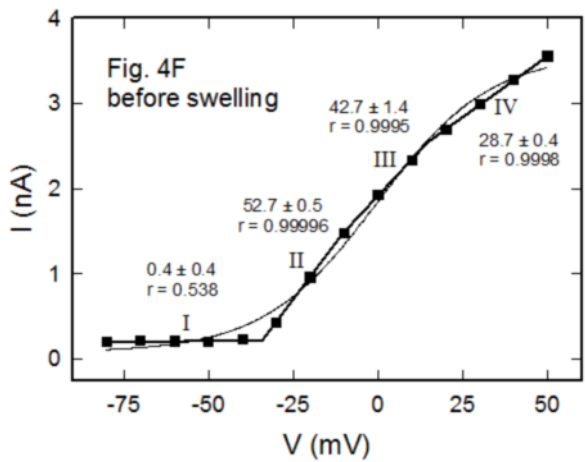

**Fig. 23.** The profile can be well represented as tetraphasic, with the transitions at -33.9, -10 and 14.9 mV. Lines II-IV have very high r values. Runs test: P = 7.75%.

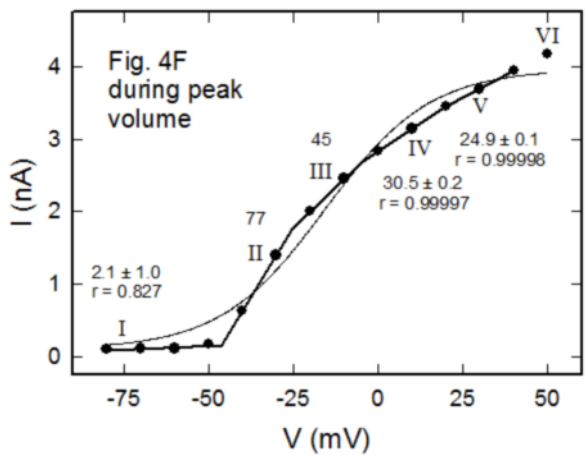

**Fig. 24.** The profile can be well represented as hexaphasic, with the transitions at -46.0, -25.3, -4.5 and 20, and between 40 and 50 mV. The r values for lines IV and V are very high. Runs test: P = 3.90%.

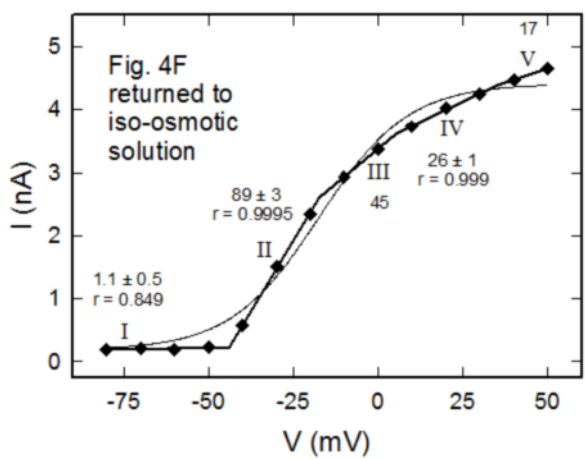

**Fig. 25.** The profile can be well represented as pentaphasic, with the transitions at -44.0, -17.3, 5.3 and 35.2 mV. The r values for lines II and IV are high. Runs test: P = 3.90%.

The probability that the uneven distribution around the three profiles in Fig. 22 is due to chance is less than 1%. Notice also how the multiphasic profiles deviate from the curvilinear profiles in about the same way in the three plots.

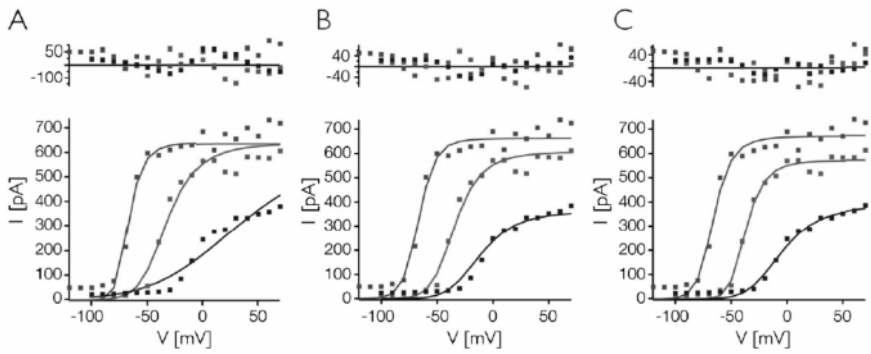

**Fig. 26.** Authors' Fig. S2 and equation 1 (right). Kv2.1 channels in the same on-cell patch with 0 mm Hg (right plot in each panel), 5 mm Hg (center plot), and 15 mm Hg (left plot). The curvilinear lines are fits to equation 1 with the following constraints: (A) $V_m$ and z independently fit for each pressure value, L constrained to be the same for all pressure values. (B) $V_m$ and z constrained to be the same for all pressure values, L independently fit for each pressure value. (C) $V_m$, z and L independently fit for each pressure value. Fitting residuals are plotted above each panel. See also the original legend.

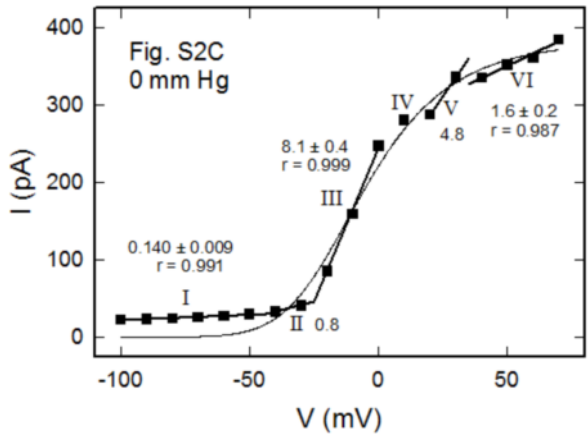

**Fig. 27.** Reanalysis of data in Fig. 26. The profile can be well represented by 7 phases, with the transitions at -43.4 and -24.7, between 0 and 10, between 10 and 20, between 30 and 40 (jump), and between 50 and 60 mV (jump). The data are insufficiently detailed in the range of phase IV for resolution of the line. Lines VI and VII could possibly be a single line, but its r value (0.987) is low compared to that for line III. Runs test: P = 0.30%.

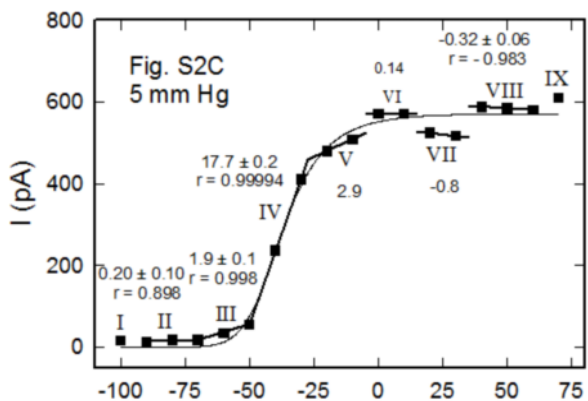

**Fig. 28.** The profile can be well represented by 9 phases, with the transitions between -100 and -90, at -70, -50 and -27.4, between -10 and 0 (jump), between 10 and 20, between 30 and 40 (jump), and between 60 and 70 mV. Line IV has a very high r value. Runs test: P = 6.00%.

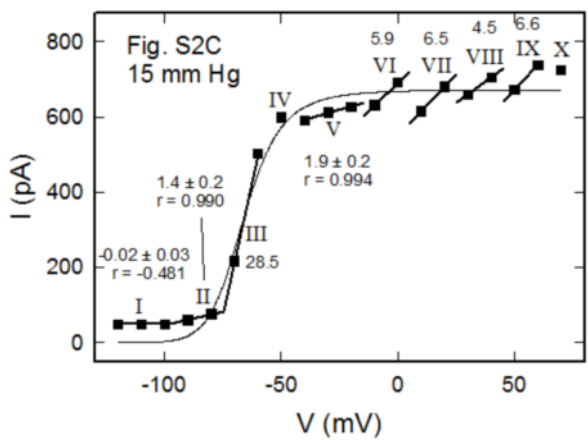

**Fig. 29.** The profile can be well represented by 10 phases, with the transitions at -100 and -74.7, between -60 and -50, between -50 and -40, between -20 and -10 (jump), between 0 and 10 mV (jump), between 20 and 30 (jump), between 40 and 50 (jump), and between 60 and 70 mV. The data are insufficiently detailed in the range of phase IV for resolution of the line. Lines VI-IX are about parallel. Runs test: P = 15.1%.

The combined probability that the uneven distribution around the three curvilinear profiles in Fig. S2A is due to chance is less than 0.5% (not shown). For Fig. S2B, the probability is less than 5% (not shown). With no constraints, i.e. for the three profiles in Fig. S2C, the combined probability is only somewhat above than 10%. (The multiphasic profiles for Figs S2A and S2B are the same as for Fig. S2C, and are not shown.) Thus, whereas the fits to the multiphasic profiles are very good, the fits to the authors' curvilinear profiles are poor even in the absence of constraints.

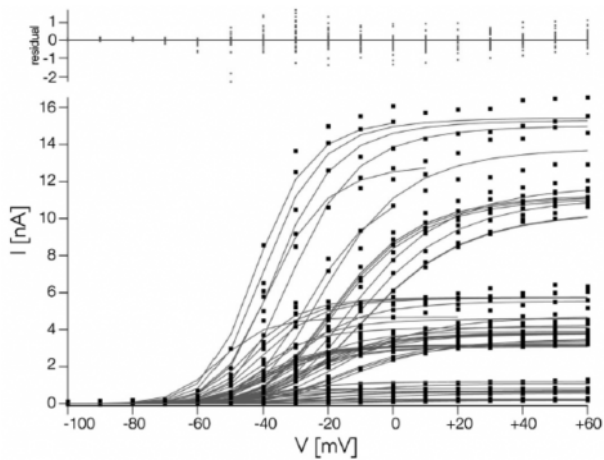

**Fig. 30.** Authors' Fig. S4. Global fit of all paddle chimera data sets. A global fit of 57 families of current extracted from 11 outside-out patches. Fitting residues are graphed above. See also original legend.

The 57 data sets in Fig. 30 have all been reanalyzed and shown to be well represented by multiphasic profiles. The profiles are not shown here, except that the probabilities for the profiles being curvilinear are included in the count below.

If the curvilinear profiles had completely represented the true situation, the experimental points should have been about equally distributed around the profiles. However, the points are, for the most part, unevenly distributed around the curves, quite often markedly so. As determined by the Runs test, an exact test, for 52 of the 75 profiles in the present analysis the probability that the uneven distribution is due to chance is less than 10%. For 43 of the profiles the probability is less than 5%, for 26 of the profiles the probability is less than 1%, and for 12 of the profiles the probability is less than 0.1%. (The probabilities for Fig. S1, reanalyzed in Nissen 2015a, are included in this count, but only the probabilities for Fig. S2C, not for Figs S2A and S2B.) The overall probability for the 75 profiles is of course vanishingly low.

# Conclusion

It is clear that the present data cannot be acceptably represented by curvilinear profiles. Thus, the points are often very unevenly distributed around the profiles, giving very low propabilities for the uneven distribution being due to chance.. The probability becomes vanishingly low if all profiles are considered together. In contrast, the data are well represented by multiphasic profiles. This conclusion also holds for other data for $K^+$ channels (Nissen 2016) as well as for a large body of other data (Nissen 2015a,b and in preparation).

**Acknowledgment –** I am very grateful to Bob Eisenberg for his continued interest and encouragement.